\documentclass[conference, 10pt]{IEEEtran}
\IEEEoverridecommandlockouts
\usepackage{cite}
\usepackage{amsmath,amssymb,amsfonts}
\usepackage{amsthm}
\usepackage{algorithmic}
\usepackage{graphicx}
\usepackage{textcomp}
\usepackage{xcolor}
\usepackage{makecell}
\usepackage{multirow} 
\usepackage{todonotes}
\usepackage{tcolorbox}
\usepackage{mathtools}
\usepackage{diagbox}
\usepackage{array}
\usepackage{verbatim}
\usepackage{bm}
\usepackage{upgreek}
\usepackage{tabularx}
\usepackage{cleveref}
\usepackage{tikz}
\usetikzlibrary{shapes.geometric, arrows}

\newtheorem{proposition}{Proposition}
\newtheorem{definition}{Definition}

\newtheorem{lemma}{Lemma}
\usepackage[caption=false,font=footnotesize]{subfig}

\theoremstyle{remark}

\def\BibTeX{{\rm B\kern-.05em{\sc i\kern-.025em b}\kern-.08em
    T\kern-.1667em\lower.7ex\hbox{E}\kern-.125emX}}

\makeatother
\makeatletter
\renewcommand{\citepunct}{,\penalty\@m\hskip.13emplus.1emminus.1em}
\renewcommand{\citedash}{\hbox{--}\penalty\@m}

\begin{document}

\title{Functional WMMSE Algorithm for Multiuser Continuous Aperture Array Systems
}
\author{
\IEEEauthorblockN{Shiyong Chen\IEEEauthorrefmark{1},
                  Shengqian Han\IEEEauthorrefmark{1},
                  Jia Guo\IEEEauthorrefmark{2}}
\IEEEauthorblockA{\IEEEauthorrefmark{1}School of Electronics and Information Engineering, Beihang University, Beijing 100191, China\\
Email: \{shiyongchen, sqhan\}@buaa.edu.cn}
\IEEEauthorblockA{\IEEEauthorrefmark{2}School of Electronic Engineering and Computer Science, Queen Mary University of London, London E1 4NS, U.K.\\
Email: jia.guo@qmul.ac.uk}
}

\addtolength{\textheight}{.1cm}
\maketitle

\begin{abstract}
In this paper, we develop a functional weighted minimum mean-squared error (WMMSE) algorithm for downlink beamforming in multiuser continuous aperture array (CAPA) systems where both the base station (BS) and users are equipped with CAPAs. We first present a closed-form expression for the achievable rate in multiuser CAPA systems, based on which the equivalence between maximizing the sum rate and minimizing the sum of weighted mean-squared errors (MSE) is established. We then employ the orthonormal basis expansion to transform the formulated functional optimization problem into a parameter optimization problem. By deriving the first-order optimality conditions of the parameter optimization problem and mapping them back to the functional domain, we obtain the update equations of the proposed functional WMMSE algorithm. Simulation results show that the proposed method outperforms discretization-based baselines in both sum rate and computational complexity.
\end{abstract}


\vspace{-.2cm}
\section{Introduction}
Continuous aperture array (CAPA) systems have emerged as a potential technique for future wireless communication, offering enhanced beamforming flexibility over conventional spatially discrete antenna arrays (SPDAs)~\cite{CAPA_based, Continuous_aperture_arrays}. A CAPA system can be viewed as an extreme SPDA where the number of antennas approaches infinity. This turns the conventional beamforming design problem into the task of determining a continuous current distribution function on the CAPA. The shift from a discrete to a continuous domain inherently leads to functional optimization problems, which cannot be solved using conventional convex optimization methods.

Recent work has investigated the optimization of beamforming for CAPA systems. A widely used approach is to discretize the continuous functions, making it suitable for conventional optimization methods. For instance, Fourier-based methods approximate the continuous beamforming and channel functions with a finite set of Fourier series~\cite{Wavenumber, Pattern_Division, On_the_SE}. This approach has been applied to single-user multi-stream CAPA systems~\cite{Wavenumber} and later extended to multiuser uplink and downlink scenarios~\cite{Pattern_Division, On_the_SE}. Through the discretization, only the coefficients of the basis functions are optimized, with which the continuous functions can be reconstructed. While this indirect optimization effectively circumvents the challenging functional optimization, it introduces approximation errors that degrade performance and necessitates a large number of basis functions, which increases with carrier frequency and CAPA size, resulting in high complexity~\cite{Beamforming_Design}.

To directly optimize continuous beamforming functions, the calculus of variations (CoV) method has been applied to CAPA systems~\cite{Beamforming_Design,Multi_Group_Multicast}. In~\cite{Multi_Group_Multicast}, an iterative CoV-based block coordinate descent algorithm was proposed for downlink multicast scenarios, where a base station (BS) equipped with CAPA serves multiple single-antenna users. In~\cite{Beamforming_Design}, a CoV-based beamforming algorithm was designed for cases where both the BS and a single user are equipped with CAPAs. However, these methods do not apply to the case where both the BS and multiple users are equipped with CAPAs. This multiuser multi-CAPA system results in a more complex functional optimization problem than those addressed in~\cite{Beamforming_Design,Multi_Group_Multicast}.

In this paper, we tackle the functional beamforming optimization for multiuser multi-CAPA systems to maximize the sum rate. We first present a closed-form expression for the achievable rate and establish the equivalence between maximizing the sum rate and minimizing the sum of weighted mean-squared errors (MSE). Then, we propose a functional Weighted Minimum Mean-Squared Error (WMMSE) algorithm to optimize the beamforming functions. By employing the orthonormal basis expansion, we obtain the first-order optimality conditions for the beamforming functions, from which the iterative update equations for WMMSE are derived. Simulation results demonstrate that the proposed algorithm outperforms discretization-based baselines in terms of both achievable sum rate and computational complexity.

\section{System Model}
\begin{figure}
\centering
 \includegraphics[width=0.3\textwidth]{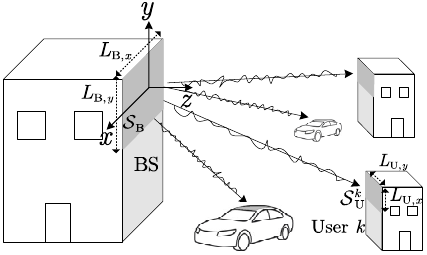}
\vspace{-0.3cm}
\caption{Illustration of the downlink CAPA system.}  \label{CAPA}
\end{figure}
Consider a downlink multiuser multi-CAPA system, where a BS equipped with a CAPA serves $K$ users, each equipped with a CAPA, as illustrated in Fig.~\ref{CAPA}. In a Cartesian coordinate system, the BS's CAPA is modeled as a rectangular aperture, lying in the $xy$-plane and centered at the origin, with side lengths $L_{\mathrm{B}}^x$ and $L_{\mathrm{B}}^y$ along the $x$- and $y$-axes, respectively. A point on the BS aperture is denoted by $\mathbf{s}\!\in\!\mathbb{R}^{3\times1}$, and we define the set of all such points as $\mathcal{S}_\mathrm{B}$. The $k$-th user's CAPA is centered at location $\mathbf{r}_{o}^k$ and has side lengths $L_{\mathrm{U}}^{x}$ and $L_{\mathrm{U}}^{y}$. Its orientation is defined by rotation angles $\omega^k_x$, $\omega^k_y$ and $\omega^k_z$ about the $x$-, $y$-, $z$-axes, respectively. The corresponding rotation matrices are $\mathbf{R}_x(\omega^k_x)$, $\mathbf{R}_y(\omega^k_y)$, and $\mathbf{R}_z(\omega^k_z)$. The set of all points in the $k$-th user CAPA is denoted by $\mathcal{S}^k_\mathrm{U}$.

The BS transmits $d$ data streams to each user. Let $\mathbf{v}_k(\mathbf{s})=[v_{1}^k(\mathbf{s}), \ldots, v_{d}^k(\mathbf{s})]\in\mathbb{C}^{1\times d}$ denote the transmit beamforming functions on the BS CAPA and $\mathbf{x}_k=[x_{1}^k, \ldots, x_{d}^k]\in\mathbb{C}^{1\times d}$ denote the data symbols for user $k$, where $v_{i}^k(\mathbf{s})$ conveys symbol $x_{i}^k$ with $\mathbb{E}\{|x_i^k|^2\} = 1$. The received signal at point $\mathbf{r}$ on the $k$-th user CAPA is given by
\begin{equation}\label{receive yk}
\!\!{y}_k(\mathbf{r})\!= \sum_{i=1}^K{\int_{\mathcal{S}_{\mathrm{B}}}{\!h_k\left( \mathbf{r},\mathbf{s} \right) \mathbf{v}_i(\mathbf{s})\mathbf{x}^{\mathsf{T}}_i\mathrm{d}\mathbf{s}+n_k(\mathbf{r})}},\, \mathbf{r}\in \mathcal{S}_{\mathrm{U}}^k,
\end{equation}
where $h_k(\mathbf{r}, \mathbf{s})$ denotes the continuous channel kernel\footnote{In functional analysis, a two-variable function used in an integral operator, such as $h_k(\mathbf{r}, \mathbf{s})$ in~\eqref{receive yk}, is referred to as the kernel of that operator.} from point $\mathbf{s}$ on the BS CAPA to point $\mathbf{r}$ on the $k$-th user CAPA, and $n_k(\mathbf{r}) \sim \mathcal{CN}(0, \sigma^2)$ is additive white Gaussian noise with variance $\sigma^2$.

CAPA is typically used in near-field communication scenarios where line-of-sight (LoS) propagation dominates. Thus, as commonly adopted in the literature~\cite{Beamforming_Design, Multi_User_CAPA}, we consider a LoS propagation model with a uni-polarized CAPA aligned along the $y$-axis, which models the channel kernel $h_k(\mathbf{r}, \mathbf{s})$ as, 
\begin{equation} \label{Green's function}
\!\!\!h_k(\mathbf{r}, \mathbf{s})\!=\!\bm{\varLambda}_{\mathrm{R}}^{\mathsf T}\frac{-j \eta e^{-j \frac{2\pi}{\lambda} \| \mathbf{r} - \mathbf{s} \|}}{2\lambda \| \mathbf{r} - \mathbf{s} \|} 
\bigg(\mathbf{I}_3\!-\!\frac{(\mathbf{r} - \mathbf{s})(\mathbf{r} - \mathbf{s})^{\mathsf T}}{\| \mathbf{r} - \mathbf{s} \|^2} \bigg)\bm{\varLambda}_{\mathrm{T}},
\end{equation}
where $\mathbf{r}\in\mathcal{S}_{\mathrm{U}}^k$, $\mathbf{s}\in \mathcal{S}_{\mathrm{B}}$, $\bm{\varLambda}_{\mathrm{T}} = [0, 1, 0]^{\mathsf T}$ and $\bm{\varLambda}_{\mathrm{R}} = \mathbf{R}_x(\omega_x^k)\mathbf{R}_y(\omega_y^k)\mathbf{R}_z(\omega_z^k)\bm{\varLambda}_{\mathrm{T}}$ are unit polarization vectors, $\eta$ denotes the intrinsic impedance of free space, $\lambda$ is the wavelength, and $\mathbf{I}_z$ denotes the $z\times z$ identity matrix.

\section{Problem Formulation}
Our objective is to optimize the beamforming functions for sum-rate maximization. Existing work has reported the closed-form expression of the achievable rate for the multiuser single-CAPA system (a CAPA-equipped BS serving single-antenna users)~\cite{Pattern_Division, Optimal_Beamforming} and the single-user multi-CAPA system (both the BS and the user are equipped with CAPA)~\cite{Beamforming_Design}. 
These results consider either intra-user interference among a user’s streams or inter-user interference across users, which therefore do not directly apply to the system we consider, where both intra-user and inter-user interference exist. 

Next, we present a closed-form expression for the achievable rate of the considered multiuser multi-CAPA system. We begin by defining the inverse of a continuous kernel.

\begin{definition}\label{Definition:Inverse of continuous kernel}
For a continuous kernel $G(\mathbf r, \mathbf s)$, a kernel $G^{-1}(\mathbf z, \mathbf r)$ is defined as the inverse of $G(\mathbf r, \mathbf s)$ if it satisfies~\cite{Optimal_Beamforming}
\begin{equation}\label{inverse of function}
 \int_{\mathcal S} G^{-1}(\mathbf z,\mathbf r)G(\mathbf r,\mathbf s)\mathrm d\mathbf r = \delta(\mathbf z-\mathbf s),
\end{equation}
for all $\mathbf{r}, \mathbf{s}, \mathbf{z}\in\mathcal{S}$, where $\delta(\cdot)$ is the Dirac delta function.
\end{definition}

\begin{proposition}\label{prop:achievable-rate}
The achievable rate of a multiuser multi-CAPA system is given by 
\begin{equation}\label{eq:theorem1 capacity}
R= \sum\nolimits_{k=1}^{K} \log\det\big(\mathbf{I}_{d} +\mathbf{Q}_k\big)
\end{equation}
where 
\begin{subequations}\label{proposition:WSR-CAPA}
\begin{align}
&\mathbf{Q}_k\!=\!
 \iint_{\mathcal{S}_{\mathrm{U}}}\!\!
\mathbf{a}_{kk}^{\mathsf{H}}(\mathbf{r}_1)\,
\mathrm{J}_{\bar{k}}^{-1}(\mathbf{r}_1,\mathbf{r}_2)\,
\mathbf{a}_{kk}(\mathbf{r}_2)\,
\mathrm{d}\mathbf{r}_2\,\mathrm{d}\mathbf{r}_1,\label{eq:theorem1 Q}\\
&\mathrm{J}_{\bar{k}}(\mathbf{r}_1,\mathbf{r}_2)\!=\! \sum\limits_{j=1,j\neq k}^K\!\! \mathbf{a}_{kj}(\mathbf{r}_1)\,\mathbf{a}_{kj}^{\mathsf{H}}(\mathbf{r}_2) 
\!+\!{\sigma^{2}}\,\delta(\mathbf{r}_1\!-\!\mathbf{r}_2)\!, \label{eq:theorem1 R}\\
& \mathbf{a}_{kj}(\mathbf{r}) 
=  \int_{\mathcal{S}_{\mathrm{B}}}\!h_k(\mathbf{r},\mathbf{s})\,\mathbf{v}_j(\mathbf{s})\,\mathrm{d}\mathbf{s}.\label{eq:theorem1 a}
\end{align}
\end{subequations}
Here, $\mathcal{S}_{\mathrm{U}}=\bigcup_{k=1}^K{\mathcal{S}_{\mathrm{U}}^k}$ and $h_k(\mathbf{r},\mathbf{s})$ is zero for $\mathbf{r}\notin\mathcal{S}_{\mathrm{U}}^k$.
\end{proposition}
\begin{IEEEproof}
The proof is omitted due to space constraints.
\end{IEEEproof}

Based on Proposition~\ref{prop:achievable-rate}, the beamforming optimization problem for maximizing the achievable rate is formulated as 
\begin{subequations}\label{P1:WSR-CAPA}
\begin{align}
\max_{\mathbf{v}_k(\mathbf{s})} 
&\sum\nolimits_{k=1}^{K} \log\det\big(\mathbf{I}_{d} +\mathbf{Q}_k\big) \\
\text{s.t.}\,\, &\sum\nolimits_{k=1}^K\int_{\mathcal{S}_{\mathrm{B}}}{\| \mathbf{v}_k(\mathbf{s}) \|^2} \mathrm{d}\mathbf{s}\le C_{\max},\label{P1:power constraint}\\
&\eqref{eq:theorem1 Q},\eqref{eq:theorem1 R},\eqref{eq:theorem1 a},\nonumber
\end{align}
\end{subequations}
where~\eqref{P1:power constraint} is the current budget constraint with $C_{\max}$ denoting the maximal current. 

To solve problem~\eqref{P1:WSR-CAPA}, we propose an iterative algorithm that is analogous to the conventional WMMSE algorithm. We first transform problem~\eqref{P1:WSR-CAPA} into an equivalent sum MSE minimization problem. Specifically, by applying the combining function $\mathbf{u}_k(\mathbf{r})\in \mathbb{C} ^{1\times d}$ to the received signal, the $k$-th user estimates its transmitted signal as
\begin{equation} \label{eq:estimated signal}
\hat{\mathbf{x}}_k= \int_{\mathcal{S}^k _{\mathrm{U}}}{\mathbf{u}_{k}^{\mathsf H}\left( \mathbf{r} \right) y_k\left( \mathbf{r} \right)}\mathrm{d}\mathbf{r}.
\end{equation}
With~\eqref{eq:estimated signal}, the MSE matrix $\mathbf{E}_k$ can be derived as 
\begin{equation}\label{eq:MSE matrix}
\begin{split}
\raisetag{7.0ex}
&\mathbf{E}_k=\mathbb{E}_{\mathbf{x},n}\!\left[
 (\hat{\mathbf{x}}_k-\mathbf{x}_k)(\hat{\mathbf{x}}_k-\mathbf{x}_k)^{\mathsf H}\right]
\\
&\triangleq\mathbf{I}_d\!-\!\mathbf{B}_{kk}\!-\!\mathbf{B}_{kk}^{\mathsf H}\!+  \sum_{j=1}^K\mathbf{B}_{kj}\mathbf{B}_{kj}^{\mathsf H}\!+\!\sigma^{2}\int_{\mathcal{S} _{\mathrm{U}}^{k}}\!\mathbf{u}_k^{\mathsf H}(\mathbf r)\mathbf{u}_k(\mathbf r)\,\mathrm d\mathbf r
\end{split}
\end{equation}
where 
\begin{equation}\label{eq:equation of B}
    \mathbf{B}_{kj}=  \int_{\mathcal{S} _{\mathrm{U}}^{k}}{\mathbf{u}_{k}^{\mathsf H}\left( \mathbf{r} \right) \mathbf{a}_{kj}(\mathbf{r})}\mathrm{d}\mathbf{r}.
\end{equation}

\begin{proposition} \label{Theorem2}
Problem~\eqref{P1:WSR-CAPA} is equivalent to the following problem, in the sense that the globally optimal solution $\mathbf{v}_k(\mathbf{s})$ is identical for both problems.
\begin{subequations}\label{P2:MMSE}
\begin{align}
\!\!&\min_{\{\mathbf{W}_k,\mathbf{u}_k(\mathbf{r}),\mathbf{v}_k(\mathbf{s})\}}   \sum\nolimits _{k=1}^K{ \mathrm{Tr}\left( \mathbf{W}_k\mathbf{E}_k \right) -\log\det \!\left( \mathbf{W}_k \right)}\label{P2:object} \\
&\quad\quad\,\,\,\,\,\mathrm{s.t.}\quad
 \eqref{eq:MSE matrix},\eqref{eq:equation of B},\eqref{eq:theorem1 a}, \eqref{P1:power constraint},\nonumber
\end{align}
\end{subequations}
where $\mathbf{W}_k\succeq\mathbf{0}$ is the weight matrix of user $k$. 
\end{proposition}
\begin{IEEEproof}
See Appendix~\ref{appendix2}.
\end{IEEEproof}

\section{Functional WMMSE Algorithm}
In this section, we propose the functional WMMSE algorithm to solve problem~\eqref{P2:MMSE}. Instead of directly optimizing the continuous functions $\mathbf{u}_k(\mathbf{r})$ and $\mathbf{v}_k(\mathbf{s})$, we expand them with two sets of orthonormal basis functions, respectively, each forming a complete (infinite-dimensional) set. This reformulation transforms the original functional optimization into an equivalent optimization over the corresponding coefficient matrices, facilitating the derivation of the functional WMMSE~algorithm.

\subsection{Reformulation via Orthonormal Basis Expansion} \label{Transformation via Orthonormal Basis Expansion}
For beamforming functions $\mathbf   {v}_k\left( \mathbf{s} \right)$, $\forall k$, let $\bm{\upbeta}\left( \mathbf{s} \right)=\left[ \beta_1\left( \mathbf{s} \right) ,\ldots ,\beta_{N_s}\left( \mathbf{s} \right) \right]\in \mathbb{C} ^{1\times N_s}$ denote the orthonormal basis functions with $N_s\to\infty$, which satisfy the following orthonormality condition  
\begin{equation} \label{eq:beta unity}
     \int_{\mathcal{S} _{\mathrm{B}}}{\bm{\upbeta }^{\mathsf{H}}\left( \mathbf{s} \right) \bm{\upbeta }\left( \mathbf{s} \right)}\mathrm{d}\mathbf{s}=\mathbf{I}_{N_s}.
\end{equation}
Then, a beamforming function $\mathbf{v}_k\left( \mathbf{s} \right)$ can be expressed as a linear combination of these basis functions, i.e., ~\cite{Fourier_series} 
\begin{equation} \label{eq:v function eq beta V}
    \mathbf{v}_k\left( \mathbf{s} \right) =\bm{\upbeta }\left( \mathbf{s} \right) \mathbf{V}_k,
\end{equation}
where $\mathbf{V}_k\in \mathbb{C} ^{N_s\times d}$ denotes the coefficient matrix, which can be obtained via the following projections 
\begin{equation} \label{eq:V eq beat v function}
\mathbf{V}_k= \int_{\mathcal{S} _{\mathrm{B}}}{\bm{\upbeta }^{\mathsf{H}}\left( \mathbf{s} \right) \mathbf{v}_k\left( \mathbf{s} \right)}\mathrm{d}\mathbf{s}.
\end{equation}

Similarly, for $\mathbf{u}_k\left( \mathbf{r} \right)$, consider another set of orthonormal basis functions, denoted as $\bm{\upalpha }_k\left( \mathbf{r} \right) =\left[ \upalpha _{1}^{k}\left( \mathbf{r} \right) ,\ldots ,\upalpha _{N_r}^{k}\left( \mathbf{r} \right) \right] \in \mathbb{C} ^{1\times N_r}$ with $N_r\!\to\!\infty$, which satisfy
\begin{equation} \label{eq:alpha unity}
     \int_{\mathcal{S} _{\mathrm{U}}^{k}}{\bm{\upalpha }_{k}^{\mathsf H}\left( \mathbf{r} \right) \bm{\upalpha }_{k}\left( \mathbf{r} \right)}\mathrm{d}\mathbf{r}=\mathbf{I}_{N_r}.
\end{equation}
Accordingly, $\mathbf{u}_k\left( \mathbf{r} \right)$ can be represented as 
\begin{equation} \label{eq:u function eq U}
    \mathbf{u}_k\left( \mathbf{r} \right) =\bm{\upalpha }_k\left( \mathbf{r} \right) \mathbf{U}_k,
\end{equation}
where $\mathbf{U}_k\in \mathbb{C} ^{N_r\times d}$ is the corresponding coefficient matrix, which can be obtained as 
\begin{equation} \label{eq:U eq u function}
\mathbf{U}_k= \int_{\mathcal{S}^k _{\mathrm{U}}}{\bm{\upalpha }_{k}^{\mathsf H}\left( \mathbf{r} \right) \mathbf{u}_k\left( \mathbf{r} \right)}\mathrm{d}\mathbf{r}.
\end{equation}

Since $\bm{\upbeta }\left( \mathbf{s} \right)$ and $\bm{\upalpha }_k\left( \mathbf{r} \right)$ form complete orthonormal sets over $\mathcal{S}_{\mathrm{B}}$ and $\mathcal{S} _{\mathrm{U}}^{k}$, respectively, the continuous channel kernel $h_k\left( \mathbf{r},\mathbf{s} \right)$ can be represented as~\cite{Beamforming_Design}
\begin{equation} \label{eq:h function eq ahb}
    h_k\left( \mathbf{r},\mathbf{s} \right) =\bm{\upalpha }_k\left( \mathbf{r} \right) \mathbf{H}_k\bm{\upbeta }^{\mathsf{H}}\left( \mathbf{s} \right), 
\end{equation}
where $\mathbf{H}_k\!\in\! \mathbb{C} ^{N_r\times N_s}$ is the channel coefficient matrix, which is obtained~as 
\begin{equation}\label{eq:H eq h function}
\mathbf{H}_k= \int_{\mathcal{S} _{\mathrm{U}}^{k}}{\int_{\mathcal{S} _{\mathrm{B}}}{\bm{\upalpha }_{k}^{\mathsf H}\left( \mathbf{r} \right) h_k\left( \mathbf{r},\mathbf{s} \right) \bm{\upbeta }\left( \mathbf{s} \right)}}\mathrm{d}\mathbf{s}\mathrm{d}\mathbf{r}.
\end{equation}

With the basis expansions for $\mathbf{u}_k(\mathbf{r})$ and $\mathbf{v}_k(\mathbf{s})$, problem~\eqref{P2:MMSE} can be reformulated as
\begin{subequations}\label{P4:discret MSE}
\begin{align}
\!\! &\min_{\mathbf{W}_k,\mathbf{U}_k,\mathbf{V}_k} \sum_{k=1}^K{\left( \mathrm{Tr}\left( \mathbf{W}_k\mathbf{E}_k \right) -\log\det \!\left( \mathbf{W}_k \right) \right)}\label{P4:objective}\\
&\,\,\,\,\,\,\,\,\,\,\text{s.t.}\quad\mathbf{E}_k=\mathbf{I}_{d}\!-\!\mathbf{B}_{kk}\!-\!\mathbf{B}_{kk}^{\mathsf{H}}\nonumber\\
&\quad\quad\quad\quad\quad\quad\quad\quad+\sum\nolimits_{j=1}^K{\mathbf{B}_{kj}\mathbf{B}_{kj}^{\mathsf{H}}}\!+\sigma^{2}\mathbf{U}_{k}^{\mathsf{H}}{\mathbf{U}}_k,\\
&\,\,\,\,\,\,\,\,\,\,\quad\quad\,\,\mathbf{B}_{kj}=\mathbf{U}_{k}^{\mathsf{H}}\,\mathbf{H}_k\,\mathbf{V}_j,\\
&\,\,\,\,\,\,\,\,\,\,\quad\quad\,\, \sum_{k=1}^K{\mathrm{Tr}\left( \mathbf{V}_{k}^{\mathsf{H}}\mathbf{V}_k \right)}\le C_{\max}.
\end{align}
\end{subequations}
Problem~\eqref{P4:discret MSE} optimizes the coefficient matrices $\mathbf{V}_k$ and $\mathbf{U}_k$ along with the weight matrix $\mathbf{W}_k$. Although it can be solved using the conventional WMMSE algorithm, the computational complexity becomes prohibitive as $N_s \!\to\!\infty$ and $N_r \!\to\!\infty$. Therefore, in the next subsection, instead of directly optimizing $\mathbf{V}_k$ and $\mathbf{U}_k$, we derive the optimality conditions for these matrices, which are then used to construct the corresponding optimal functions $\mathbf{v}_k(\mathbf{s})$ and $\mathbf{u}_k(\mathbf{r})$.

\subsection{Functional WMMSE Algorithm}
The update equations for $\mathbf{u}_k(\mathbf{r})$, $\mathbf{v}_k(\mathbf{s})$, and $\mathbf{W}_k$ are derived, respectively. 

\subsubsection{Update of $\mathbf{u}_k(\mathbf{r})$}
From problem~\eqref{P4:discret MSE}, we can obtain the first-order optimality condition for $\mathbf{U}_k$ as
\begin{equation} \label{eq:optimal condition of U}
\bigg( \sum_{j=1}^K{\mathbf{H}_k\mathbf{V}_j\mathbf{V}_{j}^{\mathsf{H}}\mathbf{H}_{k}^{\mathsf{H}}+\sigma _{}^{2}\mathbf{I}_{N_r}} \bigg) \mathbf{U}_k-\mathbf{H}_k\mathbf{V}_k=0.
\end{equation}
Multiplying both sides of~\eqref{eq:optimal condition of U} by $\bm{\upalpha }_{k}\left( \mathbf{r}_1 \right)$ yields 
\begin{equation}\label{eq:optimal condition of U1}
\!\!\bm{\upalpha}_{k}\!\left(\! \mathbf{r}_1 \!\right) \mathbf{H}_k\!\mathbf{V}_k\!=\!\bm{\upalpha}_{k}\!\left(\!\mathbf{r}_1\!\right)\!\bigg(\!  \sum_{j=1}^K\!{\mathbf{H}_k\mathbf{V}_j\mathbf{V}_{j}^{\mathsf{H}}\mathbf{H}_{k}^{\mathsf{H}}\!+\!\sigma^{2} \mathbf{I}_{N_r}} \!\bigg)\!\mathbf{U}_k.
\end{equation}
Using the orthonormality condition in~\eqref{eq:beta unity}, $\bm{\upalpha}_{k}\!\left( \mathbf{r}_1 \right) \mathbf{H}_k\!\mathbf{V}_k$ can be rewritten as 
\begin{equation} \label{eq:a function eq ahv}
\begin{aligned}
\bm{\upalpha }_{k}^{}\left( \mathbf{r}_1 \right) \mathbf{H}_k\mathbf{V}_j&= \int_{\mathcal{S} _{\mathrm{B}}}\underset{h_k\left( \mathbf{r}_1,\mathbf{s} \right)}{\underbrace{\bm{\upalpha }_{k}\left( \mathbf{r}_1 \right) \mathbf{H}_k{\bm{\upbeta }\left( \mathbf{s} \right)^{\mathsf{H}}} }}\underset{\mathbf{v}_j\left( \mathbf{s} \right)}{\underbrace{\bm{\upbeta }\left( \mathbf{s} \right)\mathbf{V}_j}}\mathrm{d}\mathbf{s}\\
&= \int_{\mathcal{S} _{\mathrm{B}}}{h_k\left( \mathbf{r}_1,\mathbf{s} \right) \mathbf{v}_j\left( \mathbf{s} \right)}\mathrm{d}\mathbf{s}\triangleq\mathbf{a}_{kj}\left( \mathbf{r}_1 \right),
\end{aligned}
\end{equation}
where the second equality comes from~\eqref{eq:h function eq ahb} and~\eqref{eq:v function eq beta V}. Substituting~\eqref{eq:a function eq ahv} into~\eqref{eq:optimal condition of U1}, we can obtain
\begin{equation}
\begin{split}\label{eq:optimal condition of U2}
\mathbf{a}_{kk}(\mathbf r_1)\!=\!\!\bigg(\! \sum_{j=1}^{K}\mathbf{a}_{kj}(\mathbf r_1)\mathbf V_j^{\mathsf H}\mathbf H_k^{\mathsf H}\!+\!\sigma^{2} \bm{\upalpha}_{k}(\mathbf r_1)\bigg)\mathbf I_{N_r}\mathbf U_k.
\end{split}
\end{equation}
With~\eqref{eq:alpha unity}, the right-hand side of~\eqref{eq:optimal condition of U2} can be further expressed~as
\begin{equation}
\begin{split}\label{eq:optimal condition of U3}
\raisetag{5.0ex}
&\int_{\mathcal S_{\mathrm U}^k}\!\!\!\bigg(\!\sum_{j=1}^{K}\mathbf{a}_{kj}(\mathbf r_1)\underset{\mathbf{a}_{kj}^{\mathsf{H}}\!\left( \mathbf{r} \right)}{\underbrace{\mathbf{V}_{j}^{\mathsf{H}}\mathbf{H}_{k}^{\mathsf{H}}\bm{\upalpha }_{k}^{\mathsf{H}}(\mathbf{r})}}\!+\!\sigma^{2} \bm{\upalpha}_{k}(\mathbf r_1) \bm{\upalpha}_{k}^{\mathsf H}(\mathbf r)\bigg)
 \bm{\upalpha}_{k}(\mathbf r)\mathbf U_k\mathrm d\mathbf r\\
&= \int_{\mathcal{S}_{\mathrm{U}}^{k}}{\bigg( \sum_{j=1}^K{\mathbf{a}_{kj}\left(\mathbf{r}_1\right) \mathbf{a}_{kj}^{\mathsf H}\left( \mathbf{r} \right) +\sigma^{2}\delta \left( \mathbf{r}_1-\mathbf{r} \right)} \bigg)\!\mathbf{u}_{k}\left( \mathbf{r} \right)}\mathrm{d}\mathbf{r},
\end{split}
\end{equation}
where $ \bm{\upalpha}_{k}(\mathbf r_1) \bm{\upalpha}_{k}^{\mathsf H}(\mathbf r)=\delta \left( \mathbf{r}_1-\mathbf{r} \right)$ holds due to the orthonormality of $ \bm{\upalpha}_{k}(\mathbf r)$, and $ \bm{\upalpha}_{k}(\mathbf r)\mathbf U_k$ equals $\mathbf{u}_{k}\!\left( \mathbf{r} \right)$ based on~\eqref{eq:u function eq U}.

Using~\eqref{eq:optimal condition of U3} to replace the right-hand side of~\eqref{eq:optimal condition of U2}, the functional optimality condition for $\mathbf{u}_k(\mathbf{r})$ is derived as
 \begin{equation}
\begin{aligned}\label{eq:functional optimal condition of U}
\!\!\!\mathbf{a}_{kk}(\mathbf r_1)\!\!=\!\! \int_{\mathcal{S}_{\mathrm{U}}^{k}}\!\!{\bigg(\! \sum_{j=1}^K{\!\mathbf{a}_{kj}\!\left(\!\mathbf{r}_1\!\right) \mathbf{a}_{kj}^{\mathsf H}\!\left( \mathbf{r} \right) \!+\!\sigma^{2}\delta \left( \mathbf{r}_1\!\!-\!\!\mathbf{r} \right)}\! \bigg)\mathbf{u}_{k}\!\left( \mathbf{r} \right)}\mathrm{d}\mathbf{r}.
\end{aligned}
\end{equation}
Defining $\mathrm{J}_{k}(\mathbf{r}_1,\mathbf{r})\triangleq \sum\nolimits_{j=1}^K{\mathbf{a}_{kj}\left(\!\mathbf{r}_1\right) \mathbf{a}_{kj}^{\mathsf H}\!\left( \mathbf{r} \right) +\sigma^{2}\delta \left( \mathbf{r}_1-\mathbf{r} \right)}$, then~\eqref{eq:functional optimal condition of U} is rewritten as $\mathbf{a}_{kk}(\mathbf r_1)= \int_{\mathcal{S}_{\mathrm{U}}^{k}}\mathrm{J}_{k}(\mathbf{r}_1,\mathbf{r}){\mathbf{u}_{k}\left( \mathbf{r} \right)}\mathrm{d}\mathbf{r}$, from which $\mathbf{u}_{k}\!\left( \mathbf{r} \right)$ can be solved as
\begin{equation}\label{eq:functional update u}
    \mathbf{u}_{k}^{}\left( \mathbf{r} \right) = \int_{\mathcal{S} _{\mathrm{U}}^{k}}{\mathrm{J}_{k}^{-1}\left( \mathbf{r},\mathbf{r}_1 \right)}\mathbf{a}_{kk}\left( \mathbf{r}_1 \right) \mathrm{d}\mathbf{r}_1,
\end{equation}
where $\mathrm{J}_{k}^{-1}\left( \mathbf{r},\mathbf{r}_1 \right)$ is the inverse of $\mathrm{J}_{k}\!\left( \mathbf{r}_1,\mathbf{r}_2 \right)$, as defined in Definition~\ref{Definition:Inverse of continuous kernel}.
\subsubsection{Update of $\mathbf{W}_k$}
From problem~\eqref{P4:discret MSE}, the optimal $\mathbf{W}_k$ can be readily obtained as
\begin{equation} \label{eq:discrete update W}
\mathbf{W}_k=\left( \mathbf{I}_{d}-\mathbf{U}_{k}^{\mathsf{H}}\mathbf{H}_k\mathbf{V}_k \right) ^{-1}.
\end{equation}
With~\eqref{eq:alpha unity},~\eqref{eq:discrete update W} can be rewritten as 
\begin{equation}
    \begin{aligned}\label{eq:functional of W}
\mathbf{W}_k&={\bigg( \mathbf{I}_d- \int_{\mathcal{S}_{\mathrm{U}}^{k}}\underset{\mathbf{u}_{k}^{\mathsf H}\left( \mathbf{r} \right)}{\underbrace{\mathbf{U}_{k}^{\mathsf H}{{\bm{\upalpha }_{k}^{\mathsf H}\left( \mathbf{r} \right)}}}} \underset{\mathbf{a}_{kk}\left( \mathbf{r} \right)}{\underbrace{\bm{\upalpha }_{k}\left( \mathbf{r} \right)\mathbf{H}_k\mathbf{V}_k}}\mathrm{d}\mathbf{r} \bigg) ^{-1}},
\\
&={\bigg( \mathbf{I}_d- \int_{\mathcal{S} _{\mathrm{U}}^{k}}\mathbf{u}_{k}^{\mathsf H}\left( \mathbf{r} \right) \mathbf{a}_{kk}\left( \mathbf{r} \right)\mathrm{d}\mathbf{r} \bigg) ^{-1}}, 
    \end{aligned}
\end{equation}
where the second equality comes from~\eqref{eq:u function eq U} and~\eqref{eq:a function eq ahv}. 

\subsubsection{Update of $\mathbf{v}_k(\mathbf{s})$}
Similar to the derivation of ${\mathbf u}_k(\mathbf r)$, the first-order optimality condition for $\mathbf{V}_k$ can be derived as
\begin{equation}\label{eq:optimal condition of V}
   \bigg(\!  \sum_{j=1}^K{\mathbf{H}_{j}^{\mathsf H}\mathbf{U}_j\mathbf{W}_j\mathbf{U}_{j}^{\mathsf H}\mathbf{H}_j}+\mu\mathbf{I} \bigg) \mathbf{V}_k-\mathbf{H}_{k}^{\mathsf H}\mathbf{U}_k\mathbf{W}_k=0,
\end{equation}
where $\mu$ is the Lagrange multiplier used to ensure the constraint in~\eqref{P1:power constraint}, which can be obtained via a bisection method. 
Multiplying both sides of~\eqref{eq:optimal condition of V} by $\bm{\upbeta }\left( \mathbf{s}_1 \right)$ yields 
\begin{equation}
\begin{aligned}\label{eq:optimal condition of V1}
\bigg(\sum_{j=1}^K\bm{\upbeta }\left( \mathbf{s}_1 \right){\mathbf{H}_{j}^{\mathsf H}\mathbf{U}_j\mathbf{W}_j\mathbf{U}_{j}^{\mathsf H}\mathbf{H}_j}+&\mu\bm{\upbeta }_{k}\left( \mathbf{s}_1 \right)\mathbf{I} \bigg) \mathbf{V}_k\\
&\quad\quad=\bm{\upbeta }\left( \mathbf{s}_1 \right)\mathbf{H}_{k}^{\mathsf H}\mathbf{U}_k.
\end{aligned}
\end{equation}
Using the orthonormality condition in~\eqref{eq:alpha unity}, the term $\bm{\upbeta }\left( \mathbf{s}_1 \right) \mathbf{H}_{k}^{\mathsf H}\mathbf{U}_k$ can be rewritten as
\begin{equation}
\begin{aligned}\label{eq:c function}
\bm{\upbeta }\left( \mathbf{s}_1 \right) \mathbf{H}_{k}^{\mathsf H}\mathbf{U}_k&= \int_{\mathcal{S}_{\mathrm{U}}^{k}}\underset{h_{k}^{\mathsf H}\left( \mathbf{r},\mathbf{s}_1 \right)}{\underbrace{\bm{\upbeta }\left( \mathbf{s}_1 \right) \mathbf{H}_k^{\mathsf H}{{\bm{\upalpha }_{k}^{\mathsf H}\left( \mathbf{r} \right)} }} }\underset{\mathbf{u}_k\left( \mathbf{r} \right)}{\underbrace{\bm{\upalpha }_{k}\left( \mathbf{r} \right)\mathbf{U}_k}}\mathrm{d}\mathbf{r}
\\
&= \int_{\mathcal{S} _{\mathrm{U}}^{k}}{h_{k}^{\mathsf H}\left( \mathbf{r},\mathbf{s}_1 \right) \mathbf{u}_k\left( \mathbf{r} \right)}\mathrm{d}\mathbf{r}\triangleq\mathbf{c}_{k}\left( \mathbf{s}_1 \right).
\end{aligned}
\end{equation}
where the second equality comes from~\eqref{eq:h function eq ahb} and~\eqref{eq:u function eq U}.
Substituting~\eqref{eq:c function} into~\eqref{eq:optimal condition of V1}, we can obtain
\begin{equation}\label{eq:optimal condition of V2}
\mathbf{c}_{k}\left( \mathbf{s}_1 \right) \mathbf{W}_k\!\!=\!\!\bigg(\! \sum\limits_{j=1}^K{\mathbf{c}_{j}\left( \mathbf{s}_1 \right) \mathbf{W}_j\mathbf{U}_{j}^{\mathsf H}\mathbf{H}_j}\!+\!\mu\bm{\upbeta }\left( \mathbf{s}_1 \right) \bigg) \mathbf{V}_k.
\end{equation}
Applying~\eqref{eq:beta unity}, the right-hand side of~\eqref{eq:optimal condition of V2} can be further derived~as 
\begin{equation}
\label{eq:optimal condition of V right hand}
\begin{split}\raisetag{2.5ex}
&\bigg(\sum_{j=1}^{K}\mathbf c_{j}(\mathbf s_1)\mathbf W_j \mathbf U_j^{\mathsf H}\mathbf H_j +\mu\,\bm\upbeta(\mathbf s_1)\bigg)\mathbf{I}_{N_s}\mathbf V_k\\
&=\int_{\mathcal S_{\mathrm B}}\bigg(\!\sum_{j=1}^{K}\!\mathbf c_{j}(\mathbf s_1)\mathbf W_j\underset{\mathbf c^{\mathsf H}_{j}(\mathbf s)}{\underbrace{\mathbf U_j^{\mathsf H}\mathbf H_j\bm\upbeta^{\mathsf H}(\mathbf s)}}+\\
&\quad\quad\quad\quad\quad\quad\quad\quad\quad\quad\quad\mu\bm\upbeta(\mathbf s_1)\bm\upbeta^{\mathsf H}(\mathbf s)\bigg)
\bm\upbeta(\mathbf s)\mathbf V_k\mathrm d\mathbf s.
\end{split}
\end{equation}
Using~\eqref{eq:optimal condition of V right hand} to replace the right-hand side of~\eqref{eq:optimal condition of V2}, the functional optimality condition for $\mathbf{v}_k(\mathbf{s})$ is derived as
\begin{equation}
\begin{aligned}
    &\mathbf{c}_{k}\left( \mathbf{s}_1 \right) \mathbf{W}_k=\\
&\,\,\, \int_{\mathcal{S}_{\mathrm{B}}}\!\!\!{\bigg(\! \sum_{j=1}^K{\!\mathbf{c}_{j}\left( \mathbf{s}_1 \right) \mathbf{W}_j\mathbf{c}_{j}^{\mathsf H}\left( \mathbf{s} \right)}\!+\!\mu\delta \left( \mathbf{s}_1\!\!-\!\mathbf{s} \right)\!\bigg)\! \mathbf{v}_k\left( \mathbf{s} \right)}\mathrm{d}\mathbf{s}.
\end{aligned}
\end{equation}
where $\bm\upbeta(\mathbf s_1)\bm\upbeta^{\mathsf H}(\mathbf s)=\delta \left( \mathbf{s}_1-\mathbf{s} \right)$ holds due to the orthonormality of $\bm{\upbeta}(\mathbf r)$, and $\bm{\upbeta}(\mathbf s)\mathbf V_k$ equals $\mathbf{v}_{k}\left( \mathbf{s} \right)$ based on~\eqref{eq:v function eq beta V}.
Defining $\mathrm{T}_k\left( \mathbf{s}_1,\mathbf{s}\right) 
\!\triangleq\!\sum_{j=1}^K{\!\mathbf{c}_{j}\!\left( \mathbf{s}_1 \right)\!\mathbf{W}_j\mathbf{c}_{j}^{\mathsf H}\left( \mathbf{s} \right)}\!+\!\mu\delta \left( \mathbf{s}_1\!-\!\mathbf{s} \right)$, the functional update equation of $\mathbf{v}_k(\mathbf{s})$ is derived~as
\begin{equation}\label{eq:functional-update-v}
    \mathbf{v}_{k}^{}\left( \mathbf{s} \right) = \!\int_{\mathcal{S} _{\mathrm{B}}}\!\!{\mathrm{T}_{k}^{-1}\!\left( \mathbf{s}_1,\mathbf{s} \right)}\mathbf{c}_{k}\left( \mathbf{s}_1 \right)\mathbf{W}_k \mathrm{d}\mathbf{s}_1.
\end{equation}
where ${\mathrm{T}_{k}^{-1}\!\left( \mathbf{s}_1,\mathbf{s} \right)}$ denotes the inverse of ${\mathrm{T}_{k}\!\left( \mathbf{s}_1,\mathbf{s} \right)}$.

The proposed functional WMMSE algorithm is summarized in Table~\ref{tab:functional-wmmse}. 

\begin{table}[t] 
  \caption{Pseudocode of the Proposed Functional WMMSE Algorithm}
  \label{tab:functional-wmmse}
  \centering
  \small
  \setlength{\tabcolsep}{6pt}
  \renewcommand{\arraystretch}{1.15}
  \begin{tabularx}{0.98\linewidth}
  {|>{\hspace{-0.5\tabcolsep}\centering\arraybackslash}p{1.0em}
    @{\hspace{3pt}}
    X<{\hspace{-0.7\tabcolsep}}|}
    \hline
    1 & \rule{0pt}{1.1em}%
        Initialize $\mathbf{v}_k(\mathbf{s})$ such that $\sum_{k=1}^K\int_{\mathcal{S}_{\mathrm{B}}}{\| \mathbf{v}_k(\mathbf{s}) \|^2} \mathrm{d}\mathbf{s}\le C_{\max}$, and set $ \mathbf W_{k}=\mathbf I_{d}$, $\forall\, k$ \\
    2 & \textbf{repeat} \\
    3 & \quad Update $\mathbf{u}_{k}(\mathbf{r})$ with~\eqref{eq:functional update u}, $\, \forall\, k$ \\
    4 & \quad Update $\mathbf W_{k}$ with~\eqref{eq:functional of W}, $\, \forall\, k$ \\
    5 & \quad Update $\mathbf v_{k}(\mathbf{s})$ with~\eqref{eq:functional-update-v}, $\, \forall\, k$ \\
    6 & \textbf{until}\ the change in $\!\sum\limits_{k=1}^K\!\log\det(\mathbf W_{k})$ is less than a tolerance~$\varepsilon$.%
        \rule[-0.2ex]{0pt}{1.1em} \\[2pt]\hline
  \end{tabularx}
\end{table}

\subsection{Implementation of the Functional WMMSE Algorithm}
Although we have obtained the closed-form update equations for $\mathbf{u}_k(\mathbf{r})$, $\mathbf{v}_k(\mathbf{s})$, and $\mathbf{W}_k$, their implementation is challenging due to the need to compute integrals at each iteration. To overcome this issue, we employ the integral approximation method~\cite{Beamforming_Design}, where the integrals over $\mathcal{S}_{\mathrm{B}}$ and $\mathcal{S}_{\mathrm{U}}^{k}$ are approximated using Gauss-Legendre quadrature. The number of sampling points for $\mathcal{S}_{\mathrm{U}}^{k}$ is denoted as $M_{\mathrm{U}}$. To maintain consistent spatial sampling density across CAPAs, the number of sampling points for $\mathcal{S}_{\mathrm{B}}$ is set to $M_{\mathrm{B}} = \big\lceil \frac{L_{\mathrm{B}}^{x}L_{\mathrm{B}}^{y}}{L_{\mathrm{U}}^{x}L_{\mathrm{U}}^{y}} \big\rceil M_{\mathrm{U}}$~\cite{Implicit_Neural_CAPA}.

In each iteration, the main complexity of the proposed functional WMMSE algorithm comes from the integral approximations, which scale with the number of sampling points as $O(M_{\mathrm{B}}^3+M_{\mathrm{U}}^3)$. In contrast, the widely used Fourier-based methods first require the computation of the coefficient matrix for $h_k(\mathbf{r},\mathbf{s})$. It has a computational complexity of $O(N_{\mathrm{B}}N_{\mathrm{U}} M_{\mathrm{B}}^2)$, where $N_{\mathrm{B}}$ and $N_{\mathrm{U}}$ represent the numbers of Fourier basis functions for the CAPAs of the BS and users, respectively, which scale quadratically with the side lengths of the surfaces $\mathcal{S}_{\mathrm{B}}$ and $\mathcal{S}^k_{\mathrm{U}}$. Moreover, computing the beamforming coefficient matrices introduces a complexity of $O(N_{\mathrm{B}}^3+N_{\mathrm{U}}^3)$~\cite{Beamforming_Design}. Consequently, the proposed algorithm is more computationally efficient than conventional Fourier-based methods.

\section{Simulation Results} \label{Simulation}
In this section, we evaluate the performance of the proposed method and compare it with relevant baselines.

The simulation setup mainly follows the configuration in~\cite{Beamforming_Design}. The number of users is $K=3$. The BS and user CAPAs have side lengths $L_{\mathrm{B}}^x=L_{\mathrm{B}}^y=2\,\text{m}$ and $L_{\mathrm{U}}^x=L_{\mathrm{U}}^y=0.5\,\text{m}$, respectively. The position of user $k$ is given by $\mathbf{r}_o^k=[r_{x}^k, r_{y}^k, r_{z}^k]^{\mathsf T}$, which is uniformly distributed with $r_{x}^k, r_{y}^k\sim U(-5,5)$ m and $r_{z}^k\sim U(20,30)$ m. The rotation angles are uniformly distributed with $\omega^k_x, \omega^k_y, \omega^k_z\sim U(-\frac{\pi}{2}, \frac{\pi}{2})$.   The wavelength is set to $\lambda=0.0107\,\text{m}$ (i.e., the carrier frequency of $28\, \text{GHz}$), and the intrinsic impedance is $\eta=120\,\pi\Omega$. To maximize the multiplexing gain, the number of data streams is set as $d=\min\{d_{\mathrm{B}}, d_{\mathrm{U}}\}$, where $d_{\mathrm{B}} = \big(2\big\lceil \frac{L_{{\mathrm{B}}}^x}{\lambda} \big\rceil + 1\big)\big(2\big\lceil \frac{L_{{\mathrm{B}}}^y}{\lambda} \big\rceil + 1\big)$ and $d_{\mathrm{U}} = \big(2\big\lceil \frac{L_{{\mathrm{U}}}^x}{\lambda} \big\rceil + 1\big)\big(2\big\lceil \frac{L_{{\mathrm{U}}}^y}{\lambda} \big\rceil + 1\big)$, as derived in~\cite{Beamforming_Design}. The maximum current budget is set to $C_{\max}=1000\,\text{mA}^2$ and the noise variance is $\sigma^2=5.6\times10^{-3}\,\text{V}^2$.   
 
We consider the following discretization-based baselines.
\begin{itemize}
    \item \textbf{Fourier}: As proposed in~\cite{Wavenumber}, the beamforming function is approximated using a finite set of Fourier series, and the finite-dimensional coefficients are iteratively optimized.
    
    \item \textbf{SPDA}: Following the discretization strategy in~\cite{Pattern_Division}, the CAPA is discretized into a finite set of spatial elements, reducing the problem to an SPDA formulation that is solved via conventional optimization methods.
\end{itemize}  

\begin{figure}[!t]
\centering
 \includegraphics[width=0.34\textwidth]{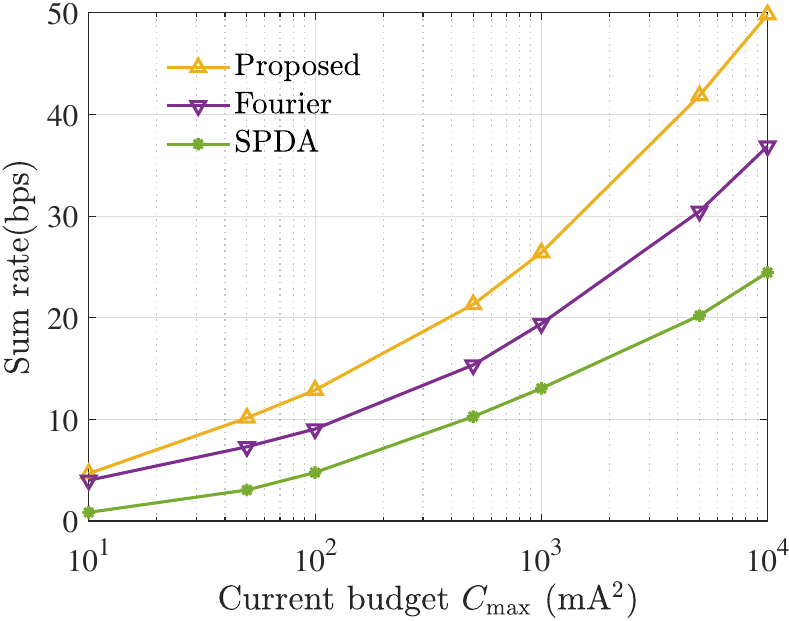}
\caption{Sum rate versus current budget.}  \label{SE_Power}
\end{figure}

Fig.~\ref{SE_Power} shows that the sum rate increases monotonically with the current budget for all schemes. The proposed method consistently outperforms the Fourier-based baseline, because the proposed method incurs only integral-approximation error, whereas the Fourier approach suffers from both discretization loss and approximation error. In addition, the proposed method yields a substantial gain over SPDA, which is attributed to the enhanced spatial degrees of freedom (DoFs) afforded by~CAPAs.

\begin{figure}[!t]
\centering
 \includegraphics[width=0.34\textwidth]{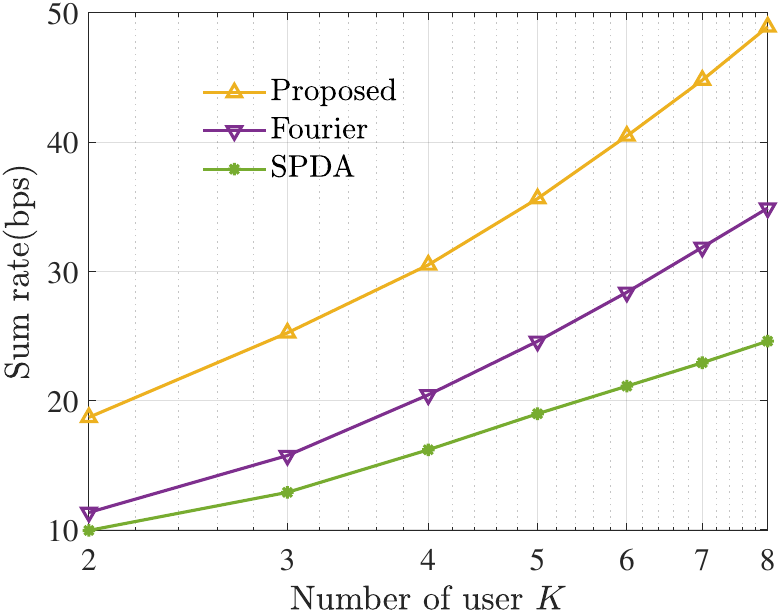}
\caption{Sum rate versus number of users.}  \label{SE_User}
\end{figure}

Fig.~\ref{SE_User} depicts the sum rate versus the number of users. The proposed approach outperforms both baselines across all user counts. This behavior reflects the stronger interference management of the functional WMMSE updates, which act directly on the continuous beamforming and combining functions to mitigate inter-user and intra-user (multi-stream) interference.

\begin{figure}[!t]
\centering
 \includegraphics[width=0.34\textwidth]{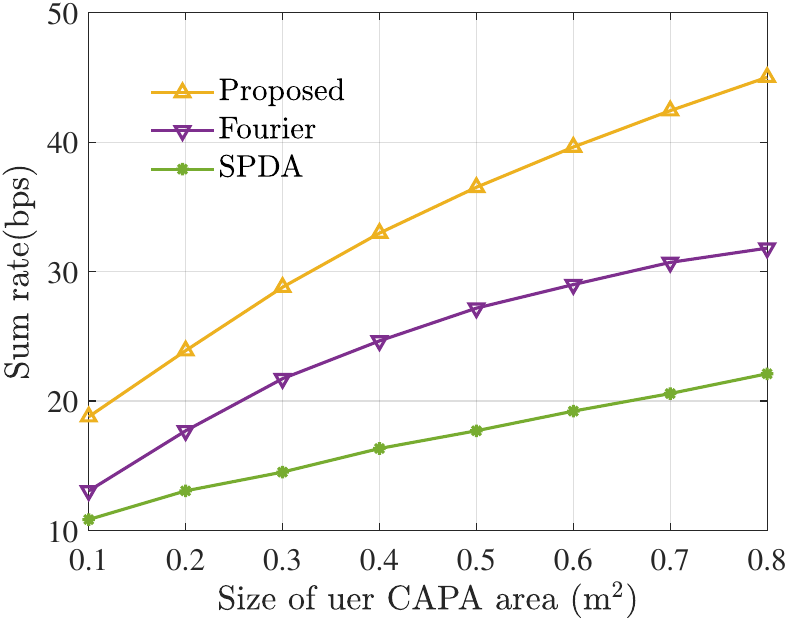}
\caption{Sum rate versus CAPA size.}  \label{SE_Size}
\vspace{-0.5cm}
\end{figure}

In Fig.~\ref{SE_Size}, we consider square user CAPAs $(L_{\mathrm{U}}^x=L_{\mathrm{U}}^y)$. Enlarging the user CAPA area improves the sum rate for all methods due to the increased spatial DoFs. The proposed approach grows faster, 
demonstrating a superior ability to exploit the DoFs enabled by larger apertures. 


Table~\ref{Complexity} compares the computational time of the proposed method with that of baselines. The proposed method significantly reduces computational time compared to the baselines.
\vspace{-1.0em}
\begin{table}[htbp]	
\centering 
\small
\caption{Comparison of Computational Time}	
\begin{tabular}{c|c|c|c}
\hline\hline
Name & \textbf{Proposed} & \textbf{Fourier} & \textbf{SPDA} \\ \hline
$A_{\mathrm{U}} = 0.25\,\mathrm{m}^2$ & 0.496 s & 7.200 s & 277.2 s\\ \hline
$A_{\mathrm{U}} = 0.5\,\mathrm{m}^2$ & 0.606 s & 13.880 s & 432.5 s  \\   \hline\hline
\end{tabular}
\begin{minipage}{0.95\linewidth}
\end{minipage}
\label{Complexity}
\vspace{-0.5em}
\end{table}

\section{Conclusions}
This paper proposed a functional WMMSE algorithm to optimize downlink multiuser beamforming for systems where both the BS and users are equipped with CAPAs. We first presented a closed-form expression for the achievable rate, with which we established the equivalence between sum-rate maximization and sum MSE minimization. Then, we transformed the functional optimization problem into the optimization of coefficient matrices through the orthonormal basis expansion. It allows us to determine the optimal conditions for the beamforming functions and subsequently derive the update equations for the functional WMMSE algorithm. Simulation results demonstrate that the proposed method achieves a higher sum rate and lower computational complexity compared to existing baselines.

\appendices
\numberwithin{equation}{section} 

\section{Proof of Proposition~\ref{Theorem2}} \label{appendix2}
Based on the first-order optimality condition of problem~\eqref{P2:MMSE} for $\mathbf{u}_{k}\left( \mathbf{r} \right)$, the optimal combining function can be readily obtained as
\begin{equation} \label{eq:optimal u}
 \mathbf{u}_{k}^{\text{opt}}\left( \mathbf{r} \right) = \textstyle\int_{\mathcal{S}^k _{\mathrm{U}}}{\mathrm{J}_{k}^{-1}\left( \mathbf{r},\mathbf{r}_1 \right) \mathbf{a}_{kk}(\mathbf{r}_1)\mathrm{d}\mathbf{r}_1},
\end{equation}
where $\mathrm{J}_{k}(\mathbf{r},\mathbf{r}_1)\!\triangleq\! \sum\nolimits_{j=1}^K\! \mathbf{a}_{kj}(\mathbf{r})\,\mathbf{a}_{kj}^{\mathsf{H}}(\mathbf{r}_1) 
\!+\!{\sigma^{2}}\,\delta(\mathbf{r}\!-\!\mathbf{r}_1)$ and $\mathrm{J}^{-1}_{k}(\mathbf{r},\mathbf{r}_1)$ denotes the inverse as defined in Definition~\ref{Definition:Inverse of continuous kernel}. Based on the first-order optimality condition for $\mathbf{W}_{k}$, we can obtain the optimal weight matrix as
\begin{equation} \label{eq:optimal W}
    \mathbf{W}_{k}^{\text{opt}}= \mathbf{E}_{k}^{-1}.
\end{equation}

Substituting $\mathbf{u}_{k}^{\text{opt}}\left( \mathbf{r} \right)$ and $\mathbf{W}_{k}^{\text{opt}}$ into~\eqref{P2:MMSE}, we can obtain the optimization problem with respect to $\mathbf{v}_k(\mathbf{s})$ as
\begin{subequations}\label{P3:WSR-CAPA}
\begin{align}
&\!\!\!\max_{\mathbf{v}_k(\mathbf{s})} 
\sum\nolimits_{k=1}^{K} \log\det\big(\big(\mathbf{E}_k^{\text{opt}}\big)^{\!-1}\big)\label{P3:objective} \\
&\text{s.t.}\quad\!\!\! \mathbf{E}_{k}^{\text{opt}}\!\!=\!\mathbf{I}_d\!-\!\!\!\iint_{\mathcal{S} _{\mathrm{U}}^k}\!\!\!\!\!\!{\mathbf{a}_{kk}^{\mathsf H}(\mathbf{r}_1)\mathrm{J}_{k}^{-1}\!\!\left( \mathbf{r}_1,\mathbf{r}_2 \right) \mathbf{a}_{kk}(\mathbf{r}_2)}\mathrm{d}\mathbf{r}_1\mathrm{d}\mathbf{r}_2,\label{P3:E opt}\\
&\,\,\,\,\,\,\,\,\,\,\eqref{eq:theorem1 a},\eqref{P1:power constraint}.\nonumber
\end{align}
\end{subequations}

\begin{lemma}\label{lem:woodbury}
Let $\mathrm{J}\left( \mathbf{r}_1,\mathbf{r}_2 \right)$ be a continuous invertible kernel and $\mathbf{a}(\mathbf{r}_1)\in \mathbb{C} ^{1\times d}$ for $\mathbf{r}_1,\mathbf{r}_2\in \mathcal{S}$. 
If the term $\mathrm{J}\left( \mathbf{r}_1,\mathbf{r}_2 \right) +\mathbf{a}(\mathbf{r}_1)\mathbf{a}^{\mathsf{H}}\left( \mathbf{r}_2 \right)$ is invertible, then
\begin{equation} \label{eq:lemma1 inverse of continuous kernel}
\begin{aligned}
&\left( \mathrm{J}\left( \mathbf{r}_1,\mathbf{r}_2 \right) +\mathbf{a}(\mathbf{r}_1)\mathbf{a}^{\mathsf{H}}\left( \mathbf{r}_2\right) \right) ^{-1}
\\
&\quad\quad\quad= \mathrm{J}^{-1}\left( \mathbf{r}_1,\mathbf{r}_2 \right) -\bm{\uppsi}(\mathbf{r}_1)\left( \mathbf{I}_d+\mathbf{G} \right)^{-1} \bm{\upphi}(\mathbf{r}_2)
,
\end{aligned}
\end{equation}
where $\mathbf{G}=  \iint_{\mathcal{S}}{\mathbf{a}^{\mathsf{H}}\left( \mathbf{r}_1 \right)  \mathrm{J}^{-1}\!\left( \mathbf{r}_1,\mathbf{r}_2 \right)\mathbf{a}(\mathbf{r}_2)}\mathrm{d}\mathbf{r}_1\mathrm{d}\mathbf{r}_2
$ and 
\begin{equation}\label{eq:lemma1 auxiliary terms}
\begin{aligned}
&\bm{\uppsi}(\mathbf{r}_1)=  \textstyle\int_{\mathcal{S}}{\mathrm{J}^{-1}\left( \mathbf{r}_1,\mathbf{r}_2 \right) \mathbf{a}(\mathbf{r}_2)\mathrm{d}\mathbf{r}_2},
\\
&\bm{\upphi}(\mathbf{r}_2)=  \textstyle\int_{\mathcal{S}}{\mathbf{a}^{\mathsf{H}}\left( \mathbf{r}_1 \right)  \mathrm{J}^{-1}\!\left( \mathbf{r}_1,\mathbf{r}_2 \right) \mathrm{d}\mathbf{r}_1}.
\end{aligned}
\end{equation}
\end{lemma}
\begin{IEEEproof}
See Appendix~\ref{appendix3}.
\end{IEEEproof}

With Lemma~\ref{lem:woodbury}, the term $\mathrm{J}_{k}^{-1}\!\!\left( \mathbf{r}_1,\mathbf{r}_2 \right)$ in~\eqref{P3:E opt} can be rewritten as
\begin{equation}\label{eq:inverse of J}
\begin{aligned}
&\mathrm{J}_{k}^{-1}\left( \mathbf{r}_1,\mathbf{r}_2 \right) =\left( \mathrm{J}_{\bar{k}}^{}\left( \mathbf{r}_1,\mathbf{r}_2 \right) +\mathbf{a}_{kk}(\mathbf{r}_1)\mathbf{a}_{kk}^{\mathsf H}\left( \mathbf{r}_2 \right) \right) ^{-1}
\\
&\quad=\mathrm{J}_{\bar{k}}^{-1}\left( \mathbf{r}_1,\mathbf{r}_2 \right) -\bm{\uppsi}_k(\mathbf{r}_1)\left( \mathbf{I}_d+\mathbf{G}_k \right)^{-1} \bm{\upphi}_k(\mathbf{r}_2),
\end{aligned}
\end{equation}
where $\mathbf{G}_k=\iint_{\mathcal{S} _{\mathrm{U}}^{k}}{\mathbf{a}_{kk}^{\mathsf H}\left( \mathbf{r}_1 \right) \mathrm{J}_{\bar{k}}^{}\left( \mathbf{r}_1,\mathbf{r}_2 \right)\mathbf{a}_{kk}(\mathbf{r}_2)}\mathrm{d}\mathbf{r}_1\mathrm{d}\mathbf{r}_2$ and the $\bm{\uppsi}_k\left( \mathbf{r}_1 \right)$ and $\bm{\upphi}_k(\mathbf{r}_2)$ are defined as
\begin{subequations}\label{eq:auxiliary terms}
\begin{align}
&\bm{\uppsi}_k\left( \mathbf{r}_1 \right) =\textstyle \int_{\mathcal{S} _{\mathrm{U}}^{k}}{\mathrm{J}_{\bar{k}}^{-1}\left( \mathbf{r}_1,\mathbf{r}_2 \right) \mathbf{a}_{kk}(\mathbf{r}_2)\mathrm{d}\mathbf{r}_2}, \\
&\bm{\upphi}_k(\mathbf{r}_2)= \textstyle\int_{\mathcal{S} _{\mathrm{U}}^{k}}{\mathbf{a}_{kk}^{\mathsf H}\left( \mathbf{r}_1 \right) \mathrm{J}_{\bar{k}}^{}\left( \mathbf{r}_1,\mathbf{r}_2 \right) \mathrm{d}\mathbf{r}_1}.
\end{align}
\end{subequations}
Substituting~\eqref{eq:inverse of J} and~\eqref{eq:auxiliary terms} into~\eqref{P3:E opt}, we can obtain
\begin{equation}\label{eq:inverse of E}
\begin{aligned}
&(\mathbf{E}_k^{\text{opt}})^{-1}=\big( \mathbf{I}_d-\mathbf{G}_k\left( \mathbf{I}_d+\mathbf{G}_k \right) ^{-1} \big) ^{-1}\!\!\!=\mathbf{I}_d+\mathbf{G}_k
\\
&\textstyle=\mathbf{I}_d+\!\! \iint_{\mathcal{S}^k_{\mathrm{U}}}{\!}\mathbf{a}_{kk}^{\mathsf{H}}(\mathbf{r}_1)\,\mathrm{J}_{\bar{k}}^{-1}(\mathbf{r}_1,\mathbf{r}_2)\,\mathbf{a}_{kk}(\mathbf{r}_2)\,\mathrm{d}\mathbf{r}_2\,\mathrm{d}\mathbf{r}_1.
\end{aligned}
\end{equation}
Combining~\eqref{eq:inverse of E} with~\eqref{P3:WSR-CAPA} completes the proof.

\section{Proof of Lemma~\ref{lem:woodbury}} \label{appendix3}
Based on Definition~\ref{Definition:Inverse of continuous kernel}, Lemma~\ref{lem:woodbury} follows from the identity
\begin{equation} \label{eq: proof of lemma1 eq1}
\begin{split}\raisetag{20.0ex}
&\int_{\mathcal{S}}\big(\big(\mathrm{J}^{-1}\left( \mathbf{r}_1,\mathbf{r} \right) -\bm{\uppsi}(\mathbf{r}_1)\left( \mathbf{I}_d+\mathbf{G} \right)^{-1}\bm{\upphi}(\mathbf{r})\big)\big.\cdot\\
&\quad\quad\big.\left(\mathrm{J}\left( \mathbf{r},\mathbf{r}_2 \right) +\mathbf{a}(\mathbf{r})\mathbf{a}^{\mathsf H}\left( \mathbf{r}_2 \right) \right)\big)\mathrm{d}\mathbf{r}\\
&\textstyle=\delta \left( \mathbf{r}_1\!-\!\mathbf{r}_2 \right)\!-\bm{\uppsi}(\mathbf{r}_1)\left( \mathbf{I}_d+\mathbf{G} \right)^{-1}\!\!\!\underset{\mathbf{G}}{\underbrace{\int_{\mathcal{S}}{\bm{\upphi}(\mathbf{r})\mathbf{a}(\mathbf{r})}\mathrm{d}\mathbf{r}}}\cdot\mathbf{a}^{\mathsf H}\left( \mathbf{r}_2 \right)-\\
&\bm{\uppsi}(\mathbf{r}_1)\!\left( \mathbf{I}_d\!+\!\mathbf{G} \right)^{\!-1}\!\!\!\!\underset{\mathbf{a}^{\mathsf H}\left( \mathbf{r}_2 \right)}{\underbrace{\int_{\mathcal{S}}\!{\!\bm{\upphi}(\mathbf{r})\mathrm{J}\left(\mathbf{r},\!\mathbf{r}_2\right)}\mathrm{d}\mathbf{r}}}\!+\!\!\underset{\bm{\uppsi}(\mathbf{r}_1)}{\underbrace{\int_{\mathcal{S}}\!{\mathrm{J}^{\!-1}\!\!\left(\mathbf{r}_1,\!\mathbf{r}\right) \mathbf{a}(\mathbf{r})}\mathrm{d}\mathbf{r}}}\cdot\mathbf{a}^{\!\mathsf H}\!\!\left(\mathbf{r}_2\right).
\end{split}
\end{equation}
Using the definitions in~\eqref{eq:lemma1 auxiliary terms}, the right-hand side of~\eqref{eq: proof of lemma1 eq1} can be simplified as
\begin{equation}
\begin{split}\raisetag{3.0ex}
&\eqref{eq: proof of lemma1 eq1}=\delta \left( \mathbf{r}_1-\mathbf{r}_2 \right) -\bm{\uppsi}(\mathbf{r}_1)\left( \mathbf{I}_d+\mathbf{G} \right)^{-1}\mathbf{G}\mathbf{a}^{\mathsf H}\left( \mathbf{r}_2 \right)\\
&\quad\quad\quad -\bm{\uppsi}(\mathbf{r}_1)\left( \mathbf{I}_d+\mathbf{G} \right)^{-1}\mathbf{a}^{\mathsf H}\left( \mathbf{r}_2 \right)+\bm{\uppsi}(\mathbf{r}_1)\mathbf{a}^{\mathsf H}\left( \mathbf{r}_2 \right)\\
&=\delta \!\left(\mathbf{r}_1\!-\!\mathbf{r}_2 \right)\!-\!\bm{\uppsi}(\mathbf{r}_1)\! \big( \!\left( \mathbf{I}_d+\mathbf{G}\right)^{\!-1}\!\mathbf{G}\!+\!\left( \mathbf{I}_d+\mathbf{G} \right)^{\!-1}\!\!-\!\mathbf{I}_d\big)\mathbf{a}^{\!\mathsf H}\!\left( \mathbf{r}_2\right)\\
&=\delta \left( \mathbf{r}_1-\mathbf{r}_2 \right),
\end{split}
\end{equation}
where $\left( \mathbf{I}_d+\mathbf{G}\right)^{-1}\mathbf{G}+\left( \mathbf{I}_d+\mathbf{G} \right)^{-1}=\mathbf{I}_d$ is used. By Definition~\ref{Definition:Inverse of continuous kernel}, the proof is completed.

\bibliographystyle{IEEEtran}
\bibliography{my_ref}
\end{document}